\documentclass[final,1p,times,authoryear]{elsarticle}
\usepackage{lipsum}

\usepackage{amssymb}
\usepackage{amsmath}
\usepackage{algorithm}
\usepackage{algpseudocode}
\usepackage{url} 
\usepackage{hyperref} 
\usepackage{textcomp}

\usepackage[dvipsnames]{xcolor}
\usepackage{comment}

\journal{Astronomy $\&$ Computing}

\graphicspath{{Figures/}}

\begin{document}

\begin{frontmatter}

\title{Assessing Performance and Porting Strategies for Gravitational $N$-Body Simulations on the RISC-V-Based Tenstorrent Wormhole\textsuperscript{\texttrademark}}

\author[1]{Jenny Lynn Almerol\corref{cor1}}
\author[1]{Elisabetta Boella\corref{cor2}}
\author[2]{Mario Spera}
\author[1]{Daniele Gregori}

\cortext[cor1]{jenny.lynn@e4company.com}
\cortext[cor2]{elisabetta.boella@e4company.com}

\affiliation[1]{organization={E4 Computer Engineering SpA},
            city={Scandiano},
            postcode={42019},
            country={Italy}}

\affiliation[2]{organization={Scuola Internazionale Superiore di Studi Avanzati (SISSA)},
            city={Trieste},
            postcode={34136},
            country={Italy}}

\begin{abstract}
% Abstract text.
While RISC-V-based accelerators were initially designed with artificial intelligence applications in mind, they are increasingly being recognized as promising platforms for high performance scientific computing. In this work, we present three strategies for scaling an $N$-body code across multiple Tenstorrent Wormhole accelerators based on the RISC-V architecture. We assess the performance of these approaches by measuring both the execution time and the energy consumption required to complete a representative simulation, ultimately identifying the configuration that offers the most favorable balance between efficiency and performance.
\end{abstract}


\end{frontmatter}


\section{Introduction}
\label{sec1}

Interest in the RISC-V Instruction Set Architecture (ISA) has grown rapidly across the High Performance Computing (HPC) community, driven by its open, royalty-free standard and its inherently modular and extensible framework~\citep{patterson_book}.
%The royalty-free, open-standard RISC-V Instruction Set (ISA) has recently gained interest in the High Performance Computing (HPC) community due to its open-source design, modularity, and extensibility~\citep{patterson_book}.
Though conceived for embedded systems, RISC-V has matured through the addition of 64-bit cores, vector processing units, and domain-specific accelerators, making it increasingly suitable for high-end computational tasks~\citep{venieri2025montecimonev2road,Teresa,MonteCimone}
%Although RISC-V was originally designed for embedded systems, recent advancements, such as 64-bit cores, vector extensions, and domain-specific accelerators, make it a promising alternative to proprietary architectures~\citep{venieri2025montecimonev2road,Teresa,MonteCimone}.
This evolution is further fueled by the rise of artificial intelligence (AI) and machine learning (ML), whose computational demands strongly overlap with those of traditional scientific workloads~\citep{brown2024accelerating,brown2025exploring}.
%These developments are further propelled by the demands of artificial intelligence (AI) and machine learning (ML), whose computational patterns align closely with many scientific workloads~\citep{brown2024accelerating,brown2025exploring}.

In scientific computing, the increasing demand for performance and energy efficiency is driving the exploration of new architectures. Astronomy and astrophysics, in particular, stand at the forefront of data-intensive science, with next-generation surveys and simulations requiring unprecedented computational throughput~\citep{SPACE1,suarez2025energy,SPACE2}. To meet these requirements, researchers are increasingly adopting heterogeneous and accelerator-based systems that can deliver high performance per Watt while scaling to massive problem sizes. Among the most computationally demanding workloads are $N$-body simulations, which model the dynamical evolution of astrophysical systems. Their reliance on large-scale linear algebra and particle–particle interactions makes them particularly well suited to hardware acceleration~\citep{khan2021adaptive, REXROTH2020100340}, and therefore ideal candidates for evaluating the potential of novel RISC-V–based accelerators.

A noteworthy example of a RISC-V-based accelerator targeting the HPC domain is the Tenstorrent Wormhole n300~\citep{wormhole}. Designed to decouple computation from data movement, it offers high parallel efficiency and energy performance at a comparatively modest cost~\citep{brown2025exploring}. 
These characteristics render it a compelling testbed for exploring the potential of RISC-V accelerators in astrophysical contexts.
%These characteristics make it an appealing testbed for evaluating the feasibility of RISC-V accelerators in astrophysical applications.

However, adopting new architectures for scientific computing presents a significant challenge, primarily due to the extensive and complex codebases developed for large-scale scientific numerical experiments, which often comprise millions of lines of code. Porting such codes is justified only when substantial performance or energy efficiency improvements can be achieved.

Motivated by this, we developed an $N$-body simulation code in C++ and offloaded the most computationally intensive calculations using the Tenstorrent TT-Metalium programming interface~\citep{metalium_guide}. To the best of our knowledge, this represents the first example of an $N$-body application ported to Tenstorrent accelerators. Thus, our application serves as a concrete test case for assessing the capability of RISC-V in demanding astrophysical workloads. The accelerated code was evaluated on the Tenstorrent Wormhole n300 card. Performance measurements show that it achieves over a $2\times$ speedup and roughly $2\times$ higher energy efficiency compared to a highly optimized baseline version executed on x86$\_$64~\citep{almerol2025SC}. 

In this work, we enhance the code parallelism by leveraging the Message Passing Interface (MPI) paradigm to use multiple accelerators simultaneously. In addition to this distributed memory approach, we implement strategies to exploit card level parallelism, enabling the utilisation of both chips present on the n300 accelerator.
As far as we are aware, this is the first scientific code capable of leveraging multiple Tenstorrent cards and multiple chips for large-scale computations. %allowing each accelerator to execute multiple computational tasks concurrently.
Thus, the  main contributions of this paper are:
\begin{itemize}
\item We explore multiple approaches to enhance the parallelization of our $N$-body code on the RISC-V-based Tenstorrent accelerators;
\item We evaluate the performance, scalability, and energy efficiency of the different proposed strategies.
\end{itemize}

%In this work, we explore the potential of RISC-V-based accelerators for astrophysical computing. We port and optimize our in-house N-body simulation code on the Tenstorrent Wormhole n300 accelerator, focusing in the computationally expensive calculations. We then evaluate its performance and energy efficiency relative to traditional CPU implementations.

\begin{comment}
Building on our preliminary report \citep{almerol2025SC}, the main contributions of this paper are as follows:
\begin{itemize}
\item We port our 
$N$-body simulation code to the RISC-V-based Tenstorrent Wormhole accelerator, exploring multiple approaches to parallelization and code adaptation;
\item We evaluate the performance, scalability, and energy efficiency of the accelerated code, and compare the results with a CPU-based reference implementation.
\end{itemize}
\end{comment}

This paper is structured as follows.
%The following sections describe our approach to exploiting parallelism on the Tenstorrent Wormhole\texttrademark{} accelerator. 
In Section \ref{sec2:background}, we introduce the direct $N$-body simulation model and the Tenstorrent Wormhole accelerator. Section \ref{sec3:porting} outlines the porting and optimization strategy. In Section \ref{sec:results} validation and benchmarking procedures are presented and performance results, including single- and multi-chip scalability as well as energy efficiency analysis, are discussed. Finally, Section \ref{sec:conclusion} illustrates concluding remarks and future perspectives.

%\section{Exploiting Parallelism on Tenstorrent Wormhole}
\section{Background} \label{sec2:background}

\subsection{Direct $N$-Body Simulation}
\label{subsec1}

Direct $N$-body simulations numerically solve the equations of motion for a system of particles by directly summing the gravitational forces exerted by all other particles in the system, expressed as:
\begin{equation} \mathbf{F}_{i} = \sum_{\substack{j=1 \\ j \neq i}}^{N} G \frac{m_i m_j}{r^3_{ij}}(\mathbf{r}_j - \mathbf{r}_i), \end{equation}
where $m_i$ and $m_j$ are the particle masses, $\mathbf{r}_i$ and $\mathbf{r}_{j}$ their position vectors, $r_{ij} = |\mathbf{r}_j - \mathbf{r}_i|$ is the inter-particle distance, and $G$ the gravitational constant.
Direct summation is the most straightforward approach to the $N$-body problem and also the most accurate, as it does not rely on approximations~\citep{spurzem1999direct}. However, this accuracy comes at a steep computational cost, since evaluating all pairwise interactions scales as $\mathcal{O}(N^2)$.
Therefore, developing \emph{accurate}, \emph{efficient}, and \emph{scalable} direct $N$-body codes is essential, particularly in advancing the astrophysical interpretation of gravitational-wave observations from current detectors such as LIGO, Virgo, and KAGRA~\citep{abbott2020prospects}, as well as future observatories such as the Einstein Telescope~\citep{einsteintelescope}.

In this work, we employ a high-order time integration scheme, the sixth-order Hermite integrator~\citep{nitadori2008sixth}, and port its implementation to the RISC-V–based Tenstorrent Wormhole accelerator.
The Hermite scheme comprises three iterative stages: prediction, evaluation, and correction~\citep{spera2014high}. In the prediction step, the positions, velocities, and accelerations of all particles are estimated from their previously known values. In the evaluation step, accelerations and their first time derivatives (jerks) are computed using these predicted quantities. Finally, in the correction step, the predicted positions and velocities are refined using the newly evaluated accelerations and jerks, achieving sixth-order time integration accuracy.

In this work, we optimize the computationally intensive all-to-all interaction of particles which involves calculating particle accelerations and jerks to exploit the parallel capabilities of the Wormhole accelerator effectively.

\subsection{The Tenstorrent Wormhole\texttrademark{} Accelerator}

\begin{figure}[t]
\centering
\includegraphics[width=0.5\linewidth]{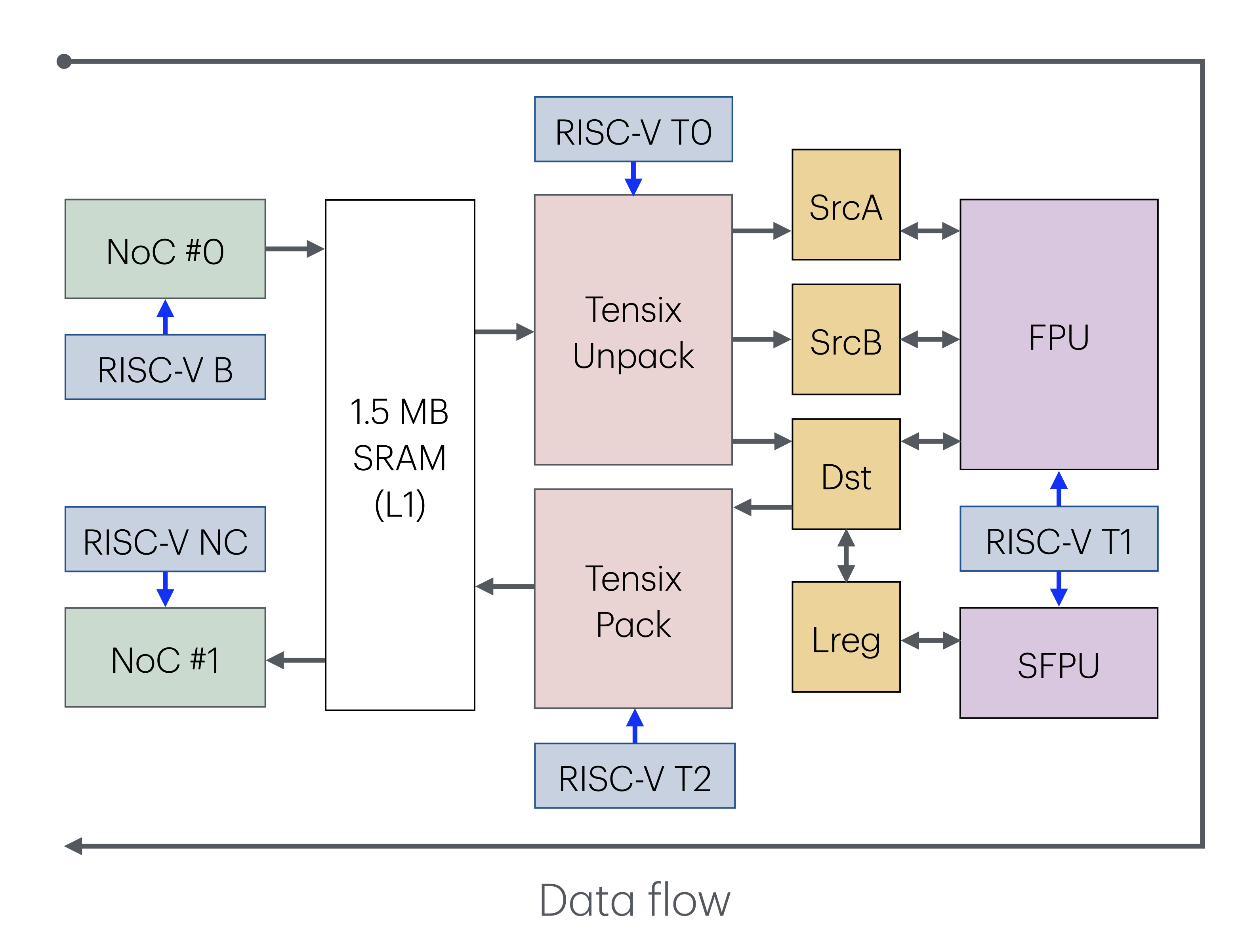}
\caption{Simplified schematic overview of a Tensix core in the Tenstorrent
Wormhole AI accelerator. Blue arrows correspond to instruction dispatch,
while black arrows indicate data movement. Adapted from~\cite{chang2025tensix}.}\label{tensix}
\end{figure}

The Wormhole is an Application-Specific Integrated Circuit (ASIC) consisting of 64 programmable processing elements, known as Tensix cores~\citep{tenstorrent_isa_docs,wormhole}. Each Tensix core contains five ``Baby'' RISC-V CPUs, 1.5~MB of local SRAM (L1), two network routers interfacing with Networks-on-Chip (NoCs), a math unit (Floating Point Unit, FPU) for low-precision matrix arithmetic supporting up to bfloat16 (BF16) operations, and a wide SIMD vector engine (Scalar Floating Point Unit, SFPU)supporting up to single-precision (FP32) operations~\citep{tenstorrent_isa_docs,corsix_tenstorrent_series}, as illustrated in Fig.~\ref{tensix}.

The ``Baby'' RISC-V processors are 32-bit, single-issue cores operating at 1~GHz~\citep{corsix_tenstorrent_series}. Two of these cores (RISC-V ``NC'' and ``B'') are dedicated to data movement, managing transfers between the Tensix core and off-chip DRAM. The remaining three are compute cores that coordinate arithmetic and logic operations by issuing instructions to the various coprocessor units, including the tensor FPU and SFPU.

Each compute core drives a specific stage of the compute pipeline~\citep{tenstorrent_isa_docs}. For instance, RISC-V~T0 (UNPACK) issues instructions to the unpacker module to load data from SRAM into two source registers, \texttt{srcA} and \texttt{srcB} for the matrix engine or the \texttt{dst} register for the vector engine. RISC-V~T1 (MATH) then issues instructions to the SFPU and FPU to perform operations on these registers. Finally, RISC-V~T2 (PACK) coordinates the movement of results from the destination register, \texttt{dst}, back to SRAM~\citep{metalium_guide}. This data flow is illustrated in Fig.~\ref{tensix}.

This hardware partitioning is mirrored in the software execution model, where we implement three custom kernels, \texttt{read}, \texttt{compute}, and \texttt{write}, each corresponding to a distinct stage of the processing pipeline. These kernels are executed across data movement and compute cores in a dataflow-driven manner, communicating via software-managed circular buffers (CBs). Kernel interactions follow a producer–consumer paradigm, with data transfers implemented as First-In, First-Out (FIFO) streams to ensure ordered and efficient communication between stages \citep{dramfifo}. 
Synchronization between kernels is managed via TT-Metalium buffer control primitives: \texttt{cb\_wait\_front}, \texttt{cb\_pop\_front}, and \texttt{cb\_reserve\_back}~\citep{kernel_api,metalium_guide}. The first two functions manage consumer-side synchronization, ensuring that a kernel proceeds only when the required input data are available and that data are consumed in the correct order. The \texttt{cb\_reserve\_back} primitive manages producer-side synchronization by suspending write operations until adequate buffer space becomes available, thereby preventing data overwrites and enforcing back-pressure. Once space is reserved, the producer writes data and invokes \texttt{cb\_push\_back} to finalize the insertion.

\section{Exploiting Parallelism on Tenstorrent Wormhole}
\label{sec3:porting}

\subsection{Porting the $N$-Body Simulation to the Tenstorrent Wormhole}

Building on the execution model described in Section~\ref{subsec1}, we port the direct $N$-body simulation to the Tenstorrent Wormhole architecture using three custom kernels: \texttt{read}, \texttt{compute}, and \texttt{write}, each responsible for a distinct stage of the computation pipeline. Data flow across these kernels is organized in a tiled layout. A tile is a 2D grid of values the architecture operates on, has a size of $32 \times 32$, and  can thus contain up to 1024 elements of the same data type. Since the Tenstorrent Wormhole natively supports up to FP32, a mixed-precision strategy is adopted for our $N$-body code. Calculations performed during the evaluation phase, including the acceleration and jerk computations on the Wormhole accelerator, are executed with FP32 precision, whereas all remaining computations on the CPU cores are performed with FP64. While in our case, this configuration is dictated by the hardware capabilities of the accelerator, previous studies have consistently shown that mixed-precision schemes deliver substantial performance and energy efficiency gains without compromising numerical accuracy~\citep{micikevicius2018mixed, schafer2020lenstool,REXROTH2020100340}.

To manage data movement across the memory hierarchy, we allocate buffers in off-chip DRAM and asynchronously copy data from the CPU to the device using non-blocking \texttt{Enqueue\-WriteBuffer} operations. The kernels then consume this data directly from DRAM. In addition, we allocate on-chip SRAM circular buffers to stage data between the \texttt{read}, \texttt{compute}, and \texttt{write} kernels.
The \texttt{read} kernel loads particle data from off-chip DRAM, which corresponds to one of the ``Baby'' RISC-V cores issuing instruction to load data to the SRAM. The \texttt{compute} kernel waits for data availability from the reader and then performs the core force calculations (see \ref{app1}). Within this kernel, the \texttt{copy\_tile} operation loads input tiles into the SFPU input register, and element-wise tile operations are executed using compute API functions such as \texttt{sub\_binary\_tile()}, \texttt{square\_tile()}, \texttt{rsqrt\_tile()}, and related primitives~\citep{kernel_api}. The results are then stored in output circular buffers via \texttt{pack\_tile}, enabling the \texttt{write} kernel to stream the computed results to the output buffers~\citep{kernel_api,metalium_guide}. 
Finally, the output data is copied back to the host using \texttt{EnqueueReadBuffer} operations.
To further optimize execution, we also implement custom ternary SFPU functions tailored to the computational patterns of the force calculation kernel, including the squared-distance calculation between particles and the \texttt{mul-add} operation needed for acceleration and jerk calculation.

Each Tensix core destination (\texttt{dst}) register can accommodate up to 8 tiles of FP32 data format. Given this limited register space, frequently reused intermediate values within the force computation, such as the displacement vector components $(r_{x,ij}, r_{y,ij}, r_{z,ij})$, are staged in on-chip SRAM using circular buffers (CBs). This staging strategy is essential for efficient data reuse and minimizing off-chip memory traffic. Consequently, the compute kernel performs multiple acquire–release operations on the \texttt{dst} registers to accommodate intermediate computations within the available register space. These acquire–release operations are necessary because the kernel must reuse more intermediate values than the 8-tile \texttt{dst} registers can hold, temporarily staging data in on-chip SRAM to avoid overwriting values needed for subsequent particle interactions.

\begin{algorithm}[t]
\caption{Asynchronous Data Loading for Particle Tiles}
\label{alg:async_load}
\begin{algorithmic}[1]
\For{each source tile $i$}
    \State \parbox[t]{.8\linewidth}{Reserve buffer space for positions $(r_x,r_y,r_z)$ and velocities $(v_x,v_y,v_z)$}
    \State \parbox[t]{.8\linewidth}{Load positions and velocities of the $i$-th particle from memory}
    \State Wait for data to become available
    \State Push loaded data to consumer buffers
    \For{each target tile $j$}
        \State \parbox[t]{.8\linewidth}{Reserve buffer space for positions $(r_x,r_y,r_z)$, velocities $(v_x,v_y,v_z)$, and weights $(p_w)$}
        \State \parbox[t]{.8\linewidth}{Load positions, velocities, and weights of the $j$-th particle from memory}
        \State Wait for data to become available
        \State Push loaded data to consumer buffers
    \EndFor
\EndFor
\end{algorithmic}
\end{algorithm}

\begin{figure}[t]
\centering
\includegraphics[width=0.5\linewidth]{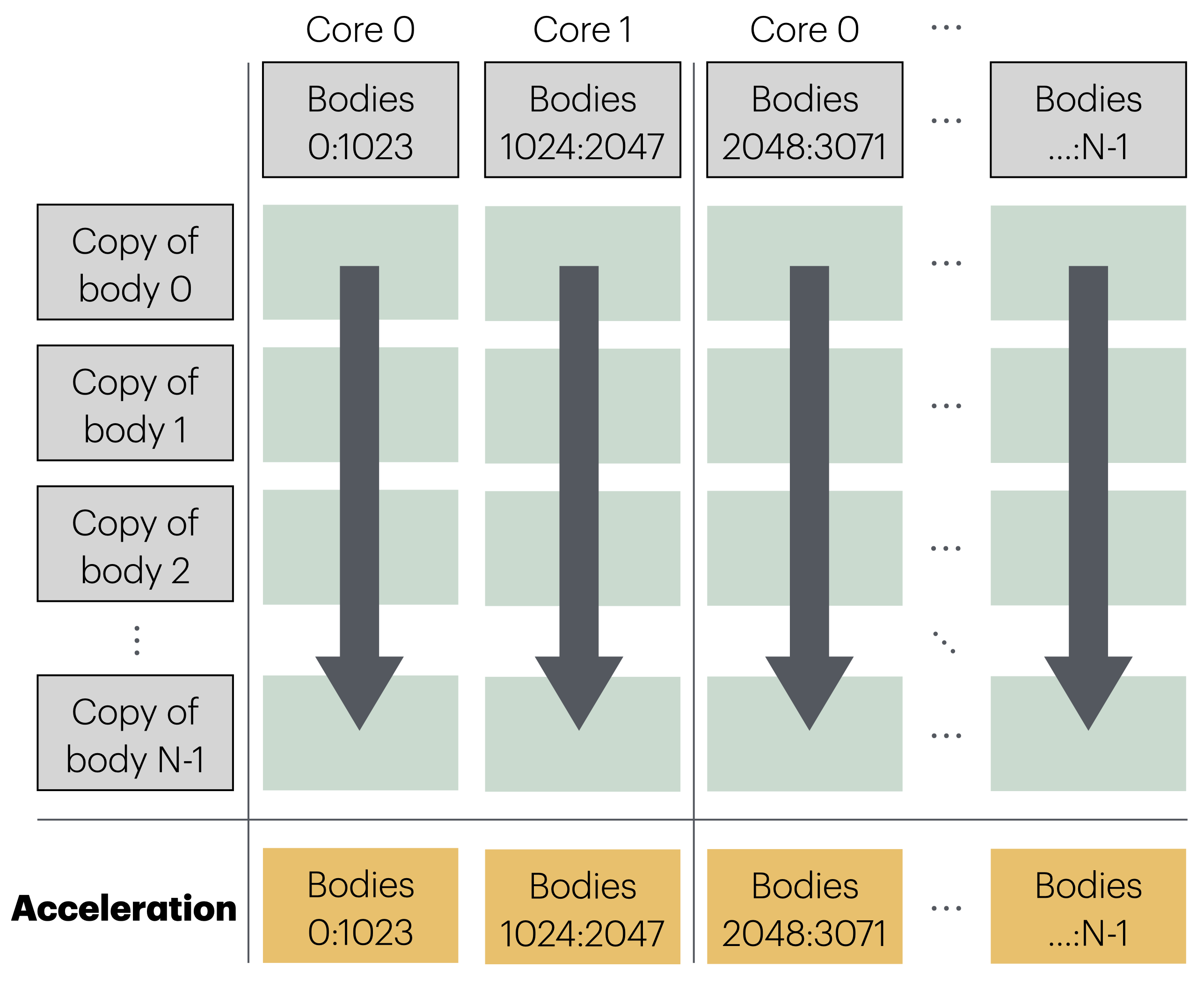}
\includegraphics[width=0.5\linewidth]{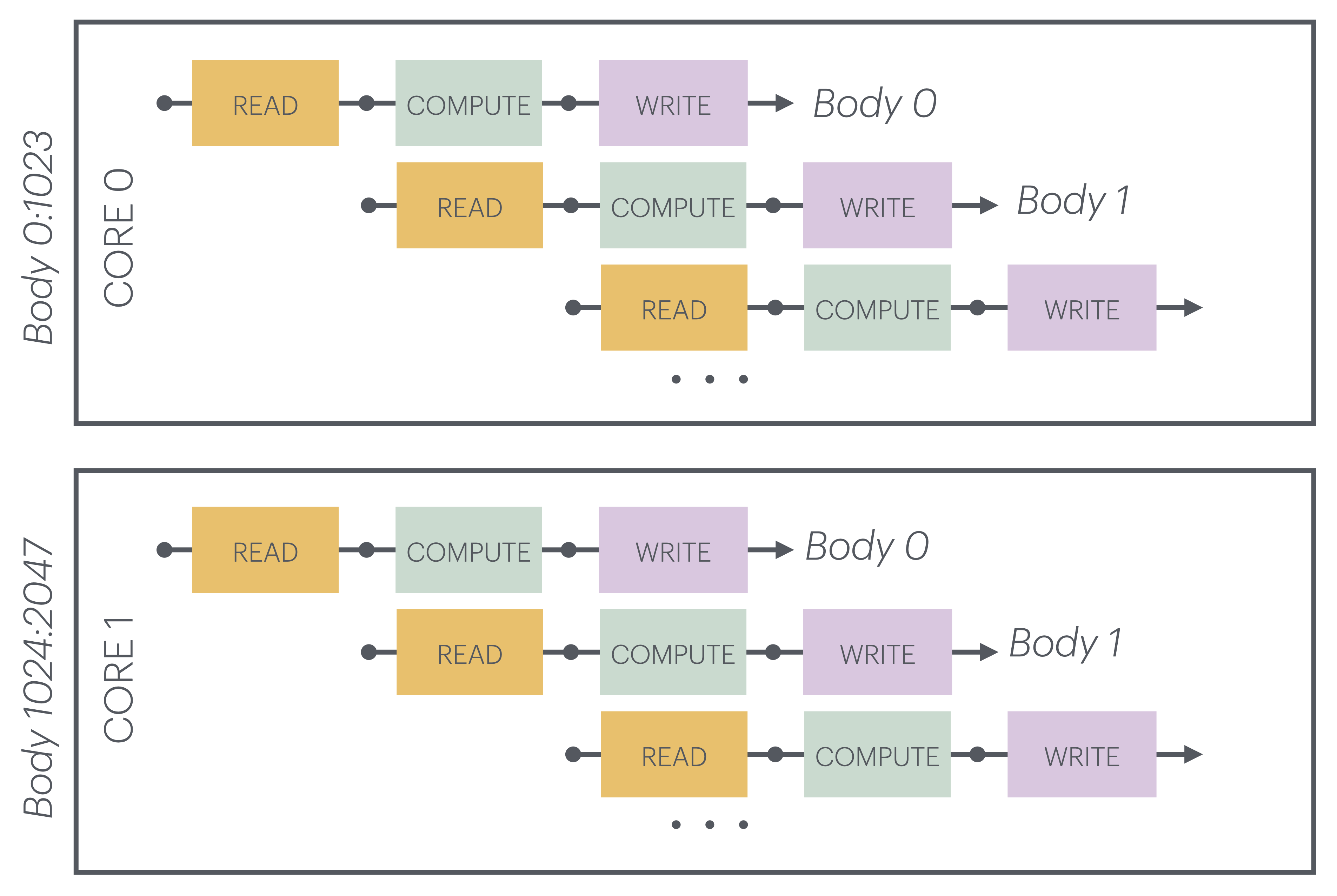}
\caption{(Top) Tile-based parallel force calculation. Gray row tiles correspond to the replicated data. Tiles within the vertical lines show the first batch of calculations when, for example, 2 Tensix cores are used. When using more Tensix cores and/or using more Wormhole devices, more tiles are processed in parallel. (Bottom) Pipelined kernel execution illustrating the overlap between data movement and computation.}\label{tiling}
\end{figure}

To enable parallel execution, we employ a Single Program, Multiple Data (SPMD) model. Each Tensix core executes the same program but operates on a distinct subset of the dataset to compute particle accelerations and jerks. Workload distribution is managed using Metalium \texttt{split\_work\_to\_cores} function~\citep{metalium_guide}. However, to compute the net gravitational force on a particle, each core must access the complete set of particle data. To accommodate this and to match the architecture tile-based data-access pattern, particle data are replicated and organized into tiles of 1024 elements, with each tile containing the same information for a single particle attribute (e.g., position or velocity component). In other words, each tile contains 1024 identical copies of one scalar quantity for a given particle. This replication strategy scales with adding more computational resources such as using multiple Wormhole cards and chips, since each chip stores its own copy of the replicated data in local DRAM. The data loading process implemented in the \texttt{read} kernel is illustrated in Algorithm~\ref{alg:async_load}, while the data-parallel organization is illustrated in Fig.~\ref{tiling}: column tiles are distributed across Tensix cores, while rows correspond to parallel computations. Acceleration and jerk values are accumulated along the indicated directions and, once complete, written to the output CBs. Fig.~\ref{tiling} also shows the pipelined execution of \texttt{read}, \texttt{compute}, and \texttt{write} kernels, allowing data movement to overlap with computation. The data consumption and computation process implemented in Algorithm~\ref{alg:async_compute} follows the same asynchronous streaming pattern as the data production stage shown in Algorithm~\ref{alg:async_load}.

\begin{algorithm}[t]
\caption{Asynchronous $N$-Body Acceleration and Jerk Computation and Accumulation}
\label{alg:async_compute}
\begin{algorithmic}[1]
\For{each source tile $i$}
    \State \parbox[t]{.8\linewidth}{Wait for input positions $(r_x, r_y, r_z)$ and velocities $(v_x, v_y, v_z)$ of the $i$-th particle}
    \State \parbox[t]{.8\linewidth}{Reserve buffer space for acceleration and jerk}
    \State \parbox[t]{.8\linewidth}{ Initialize local acceleration $(a_x^t, a_y^t, a_z^t)$ and jerk $(\dot{a}_x^t, \dot{a}_y^t, \dot{a}_z^t)$ tiles to zero}
    \For{each target tile $j$}
        \State \parbox[t]{.8\linewidth}{Wait for input positions $(r_x, r_y, r_z)$, velocities $(v_x, v_y, v_z)$ and weights $(p_w)$ of the $j$-th particle}
        \State \parbox[t]{.8\linewidth}{Compute and update pairwise accelerations and jerks (see~\ref{app1})}
        % \State \parbox[t]{.8\linewidth}{Update temporary accelerations: $\mathbf{a}^t_i$}
        % \State \parbox[t]{.8\linewidth}{Update temporary jerk: $\dot{\mathbf{a}}^t_i$}
    \EndFor
    \State \parbox[t]{.8\linewidth}{Pack accumulated $(a_x^t, a_y^t, a_z^t)$ and $(\dot{a}_x^t, \dot{a}_y^t, \dot{a}_z^t)$ into output buffers}
    \State \parbox[t]{.8\linewidth}{Push finalized acceleration and jerk to consumer buffers}
\EndFor
\end{algorithmic}
\end{algorithm}

An alternative approach to parallel execution on Wormhole devices is the \textit{mesh programming model}~\citep{meshdevice}. In this model, a \texttt{MeshDevice} acts as a virtual abstraction that unifies multiple physical devices into a single logical mesh, in principle allowing efficient distribution of computations across a multi-device topology. When operations are issued to a \texttt{MeshDevice}, command queues automatically distribute workloads to all participating devices concurrently, ideally reducing dispatch overhead compared to sequential, device-by-device execution.

To scale the simulation across multiple devices, we utilize up to four Wormhole n300 cards in isolated configuration (e.g. without any internal interconnects between cards). Each card hosts two chips: the L-chip, which connects directly to the host via PCIe, and the R-chip, which connects to the L-chip over an Ethernet link~\citep{metalium_guide}.
%connected via Ethernet where the chip with direct PCIe host connectivity (L-chip) communicates with the secondary chip (R-chip) through this Ethernet link~\citep{metalium_guide}.
The simulation is deployed in a multi-host, distributed configuration using \texttt{tt-run}, the TT-Metalium distributed process launcher, which provides a declarative YAML-based interface for MPI-based orchestration~\citep{meshdevice}. During the evaluation step, MPI is used to distribute the particle data across the Wormhole devices. Furthermore, OpenMP is employed to parallelize both the prediction and correction steps of the $N$-body calculation within each MPI process, improving overall scalability and computational throughput. The full implementation of the code can be found in the Github repository\footnote{\url{https://github.com/jlalmerol/N-Body-Code.git}}.

\paragraph{Scaling Configurations for Multi-Card Execution} As shown in Fig.~\ref{conf}, we consider three approaches to scale the $N$-body simulation across multiple Wormhole cards and chips:

\begin{figure*}[t!]
\centering
\includegraphics[width=0.3\linewidth]{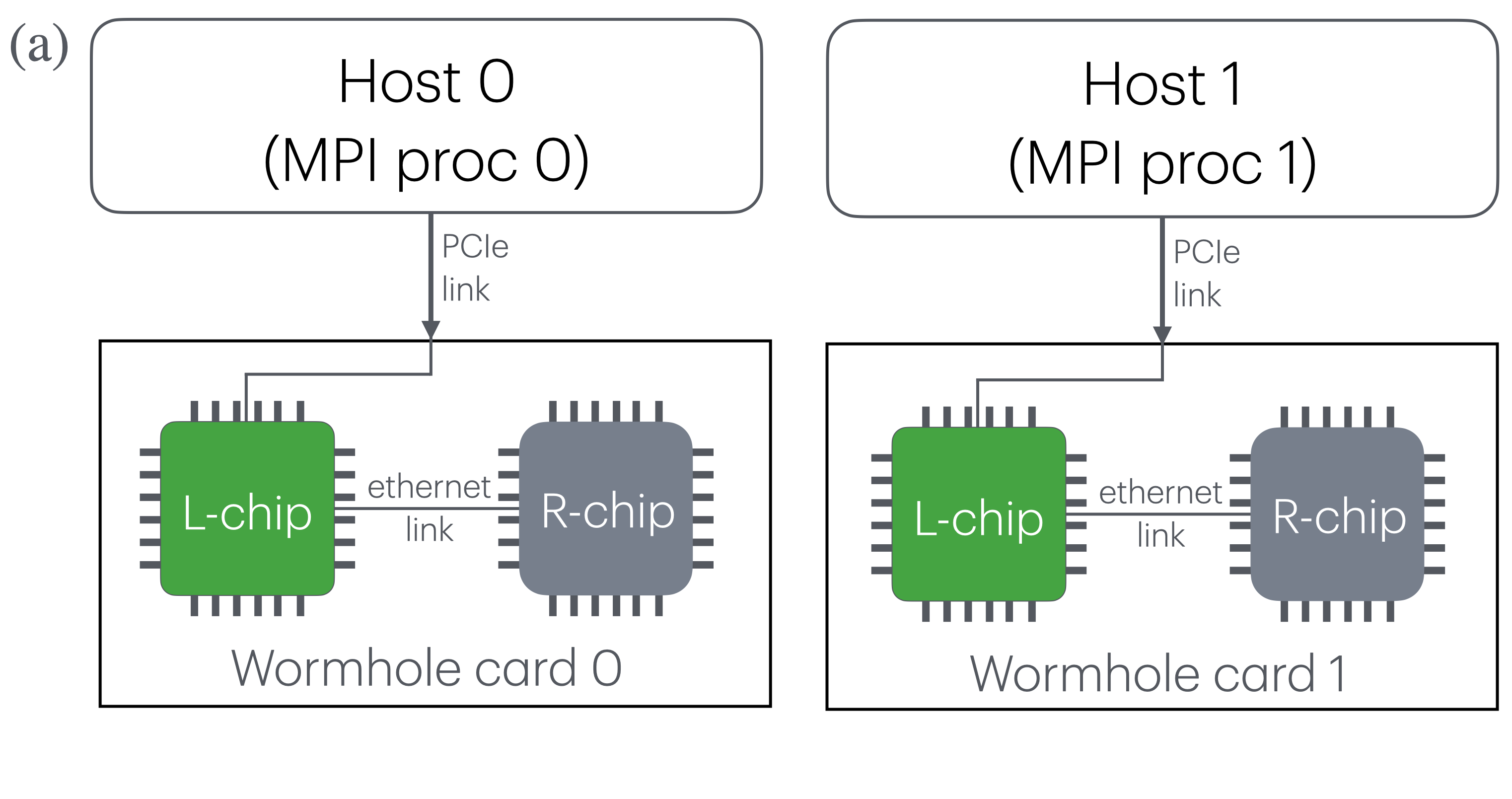}
\includegraphics[width=0.3\linewidth]{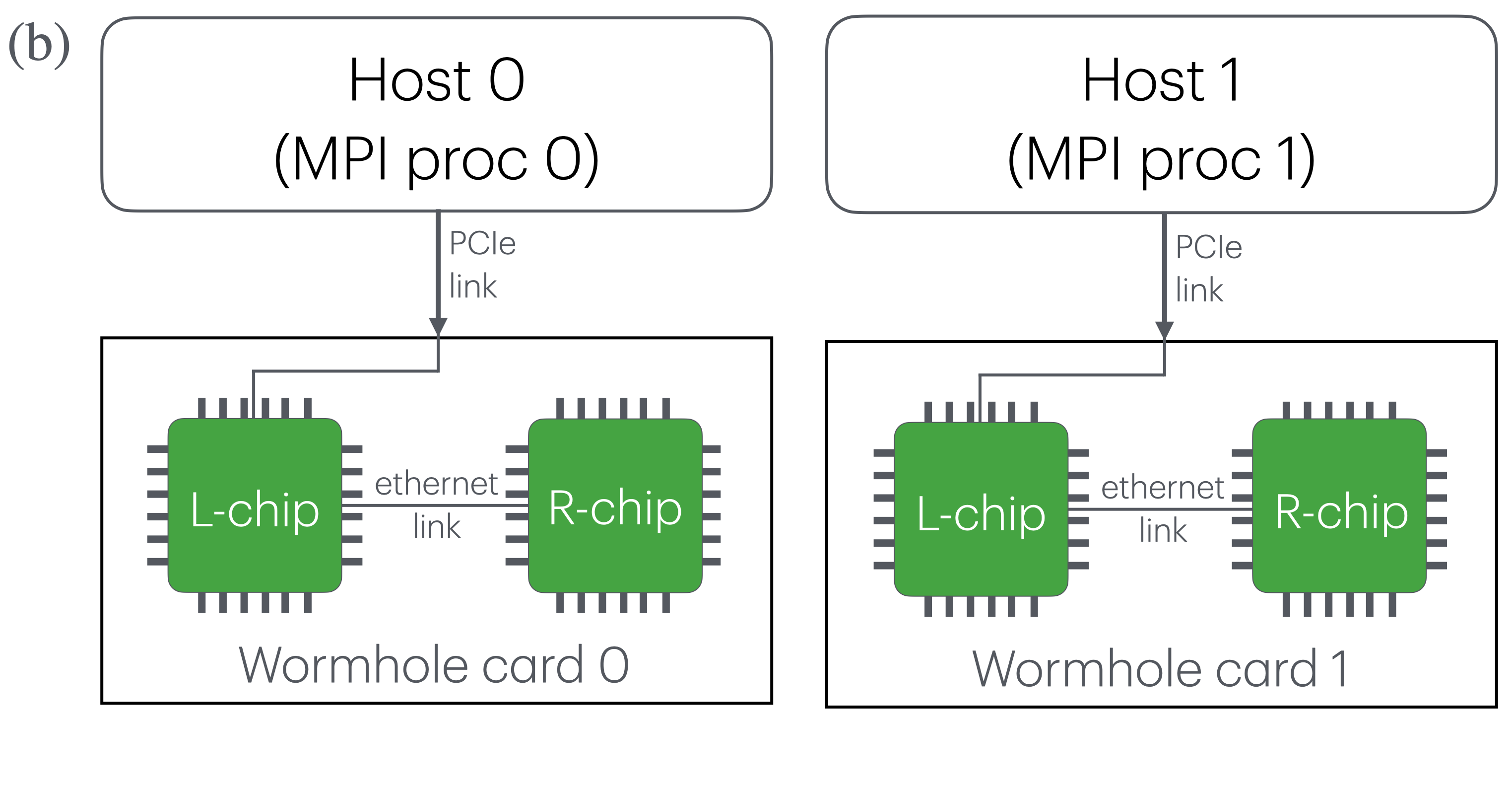}
\includegraphics[width=0.3\linewidth]{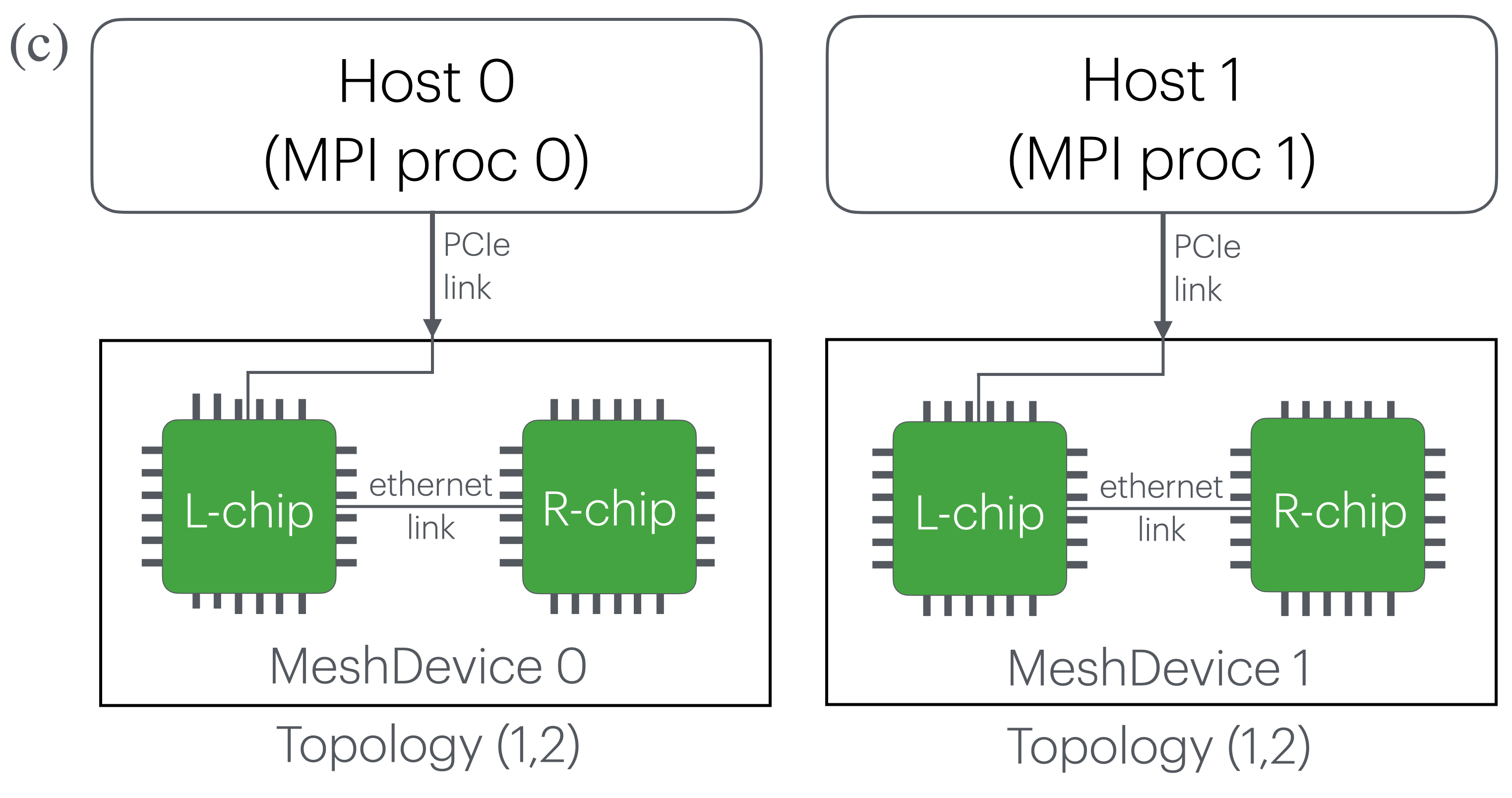}
\caption{Schematic of three parallel execution strategies for leveraging Multi-Host and Multi-Chip Wormhole architectures, illustrated for the case of two Wormhole cards: (a) Multi-Host and Single-Chip configuration, where only one chip per card is utilized; (b) Multi-Host and Multi-Chip configuration, where both chips per card are used; and (c) Mesh-Based configuration, where both chips are organized as \texttt{MeshDevice} instances.}
\label{conf}
\end{figure*}

\begin{enumerate}

\item Multi-Host and Single-Chip Configuration:
In this configuration, only the L-chip directly connected to the host via PCIe is utilized from each Wormhole card, while the corresponding R-chip remains idle. Parallel execution is achieved by assigning each active chip to a dedicated MPI process. Each MPI process independently initializes its device using the \texttt{CreateDevices} function and allocates interleaved buffers to store the subset of particle data assigned to that process. Global data decomposition is performed using MPI, and the computational workload within each chip is further distributed across Tensix cores using the \texttt{split\_work\_to\_cores} function.

\item Multi-Host and Multi-Chip Configuration:
This configuration extends the Multi-Host approach by employing both chips per card within a single MPI process. Separate program instances are defined for each device, including explicit buffer creation and program enqueueing. As a result, each MPI process iterates over its local devices to manage data transfers and kernel execution. This includes explicit data partitioning between the two chips as well as further distribution of the workload across Tensix cores using the \texttt{split\_work\_to\_cores} function.
% The workload is also distributed using \texttt{split\_work\_to\_cores}.

\item Mesh-Based Configuration:
In this configuration, parallel execution across multiple chips is managed by employing the \texttt{MeshDevice} abstraction. Unlike the previous approaches, command queues automatically distribute operations to all devices in the mesh, reducing per-device dispatch overhead and avoiding redundant per-device compilation. In our implementation, a $(1,2)$ mesh topology is instantiated per MPI process, corresponding to the two chips available on each Wormhole card. Data exchange between devices is managed through \textit{sharded buffers} for domain-decomposed data and \textit{replicated buffers} for globally shared particle data. In addition to data decomposition across Wormhole cards using MPI, sharded buffers distribute data across the devices within each card (i.e., the two chips), such that each device operates independently on its assigned subset. Replicated buffers, on the other hand, maintain a complete copy of the data on every device, ensuring that globally shared values are immediately accessible without requiring inter-device communication.

\end{enumerate}

\section{Benchmarking of the Wormhole $N$-Body Application} \label{sec:results}

%The Wormhole-based \textit{N}-body simulations are benchmarked against a CPU-based reference implementation. The reference code performs the evaluation step entirely on the CPU in parallel using MPI and OpenMP, and employs mixed precision. Additionally, the acceleration and jerk computation are further optimized leveraging AVX-512 intrinsics, maximizing single-core performance.

\subsection{Validation}
\begin{figure}[t]
\centering
\includegraphics[width=0.5\linewidth]{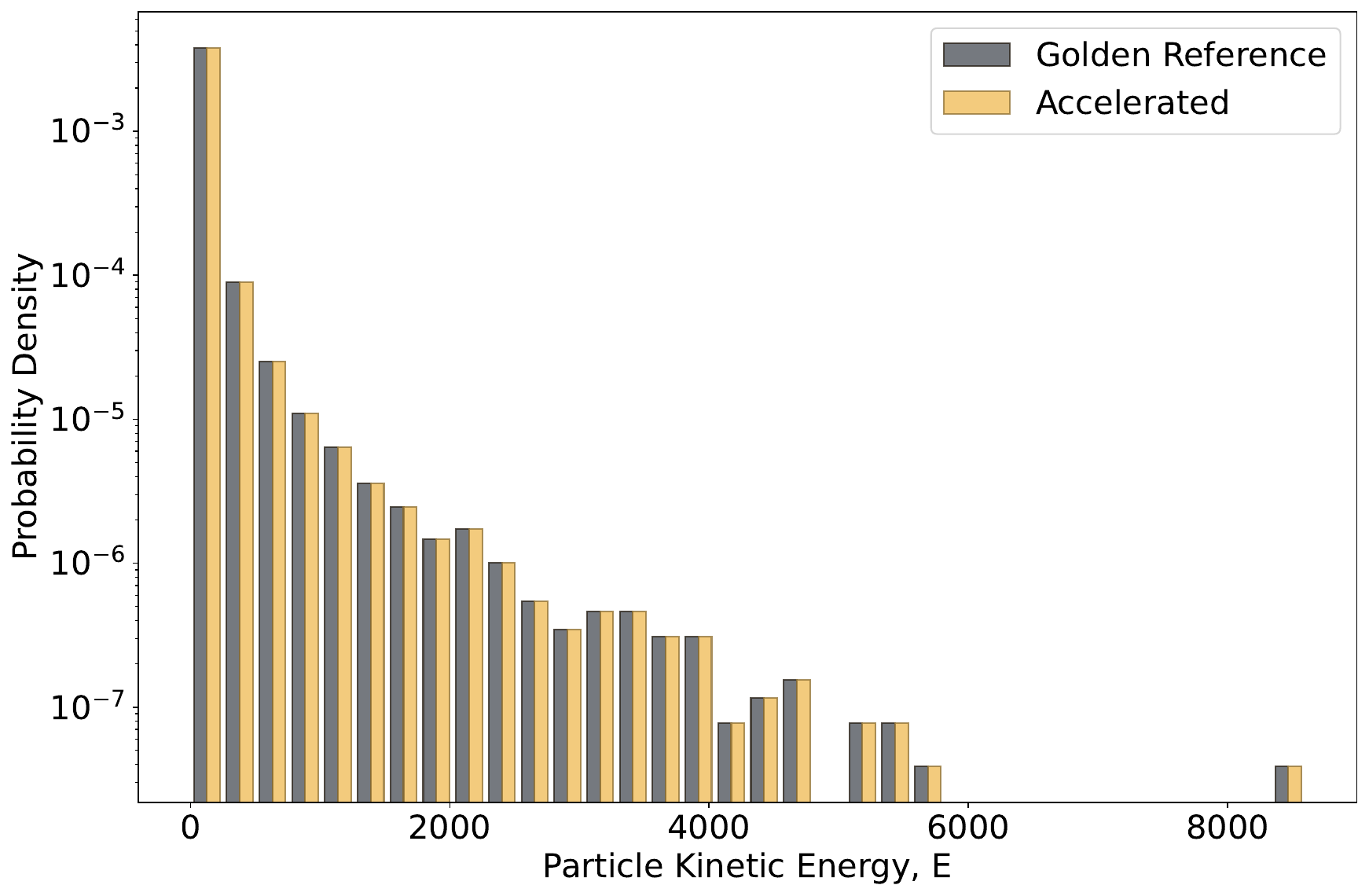}
\caption{Energy distribution of particles obtained from the Wormhole-accelerated simulation (orange) and the golden reference (gray) after time cycle $t=3$. The two distributions illustrate the consistency between the accelerated and the golden reference dataset. All quantities are expressed in normalized units.}\label{energydist}
\end{figure}

To verify the accuracy of the ported $N$-body application, the calculated acceleration and jerk values are compared against a ``golden reference'', namely the naive, double-precision brute-force implementation of the $N$-body algorithm executed serially on a conventional CPU. Discrepancies are confirmed to be within acceptable tolerances with each acceleration and jerk component deviating by no more than $0.05 \%$ and $0.2 \%$, respectively, relative to the golden reference, thereby validating the correctness of the TT-Metalium implementation and the integrity of its hardware operations. A comparison between the “golden reference” and the accelerated codes is shown in Fig. \ref{energydist}, where the energy distributions of the particles at the end of a typical simulation are reported for both cases. The accelerated simulation employs Approach~1; however, comparable distributions are observed when using the other two approaches.

\subsection{Performance evaluation}
\paragraph{Setup} For performance evaluation, the Wormhole-based $N$-body simulations are benchmarked against an optimized CPU-based reference implementation. The reference code performs the evaluation step entirely on the CPU in parallel using MPI and OpenMP, and adopts mixed-precision arithmetic to ensure consistency with the accelerated implementation. Additionally, the computation of acceleration and jerk is further optimized through the use of AVX-512 intrinsics, thereby maximizing single-core performance. We note that a detailed evaluation of the efficiency of the mixed-precision approach relative to a full double-precision implementation is beyond the scope of this work.

The reference code is compiled using CMake with the GNU Compiler Collection (GCC) version~13.3.0, conforming to the C++20 standard. Distributed parallelism is enabled through Open~MPI~version~4.0.6. The build process applies several optimization and performance-oriented flags, including \verb|-O3|, \verb|-fopenmp|, \verb|-march=native|, and \verb|-mavx512f|. 
%The same software environment is employed for the accelerated version, with the exception of Open MPI~5.0.7-ULFM, which is used in place of version~4.0.6. Additionally the software stack leverages the TT-Metal library (v0.62.2). The code is compiled with the flags \verb|-O3|, \verb|-fopenmp|, and \verb|-march=x86-64-v3|. \textcolor{blue}{We note that our code does not employ any MPI fault-tolerance features, but Open MPI 5.0.7-ULFM is a required dependency of the TT-Metal library v0.62.2 and must be installed to complete the setup of the Tenstorrent Application Programming Interface needed to operate the Wormhole cards.}
The accelerated version uses the same software environment, except that Open MPI 5.0.7-ULFM replaces version 4.0.6. This version is required as a dependency of the TT-Metal library (v0.62.2), the Tenstorrent Software Development Kit used to operate and program the cards, although our application does not employ any MPI fault-tolerance features. The accelerated code is compiled with the flags \verb|-O3|, \verb|-fopenmp|, and \verb|-march=x86-64-v3|.

%The reference code is compiled using CMake with the GNU Compiler Collection (GCC) version 13.3.0, conforming to the C++20 standard. Distributed parallelism is enabled through Open MPI version 4.0.6. The build process applies several optimization and performance-oriented flags, including \verb|-O3|, \verb|-fopenmp|, \verb|-march=native|, and \verb|-mavx512f|.
%The same software environment is employed for the accelerated version, which additionally integrates the TT-Metal library (v0.62.2), and is compiled with the flags \verb|-O3|, \verb|-fopenmp|, and \verb|-march=x86-64-v3|.

The host system used to run the $N$-body simulations for the experimental campaign is equipped with the dual-socket AMD EPYC 9124 processor, offering a total of 64 hardware threads (2 sockets $\times$ 16 cores $\times$ 2 threads per core) and a maximum clock frequency of $3.71 \, \mathrm{GHz}$. The system is equipped with $1.5 \, \mathrm{TB}$ of DDR5 memory and runs Ubuntu 24.04.2 LTS (64-bit) with Linux kernel version 6.8.0. Four Wormhole n300 devices are connected to the processor via PCIe Gen 4.

%\subsection{Results and Scalability}

\paragraph{Performance results} We evaluate the three proposed approaches by measuring the time required to complete a representative simulation (time-to-solution, in seconds) and the energy consumed by the processor and accelerators during its execution (energy-to-solution, in Joule). In this study, we consider simulations modeling 409600 particles evolving over three time steps. This low number of time steps has been chosen to keep the total execution time manageable. While this restricts the total number of simulated time steps, it does not affect the validity of the performance characterization, since the initialization phase of our $N$-body code represents only a minor fraction of the overall execution time and does not dominate the measured behavior.

We execute multiple simulations (typically 20) to obtain sufficient statistical data, launching them in batches. To ensure the system returns to idle conditions between successive runs, a 120-second \verb|sleep| period is inserted both before and after each simulation.

The time-to-solution is measured using hardcoded \verb|MPI_Wtime()| calls at the beginning and end of the simulation, excluding any time spent in \verb|sleep|. Table~\ref{tab:times} reports the average time-to-solution and the corresponding standard deviation for simulations performed using the three approaches described in Section~\ref{sec3:porting}. For reference, in the case of the Multi-Host Single-Chip scenario, we also include the execution time obtained when employing two chips (and thus two cards). All the simulations are launched using \verb|tt-run|, with each MPI task bound to one device. Additionally the Open MPI flag \verb|--bind-to core| is used and \verb|OMP_NUM_THREADS| is set to one.
In other words, we use one MPI rank per n300 card, with each MPI rank pinned to a dedicated physical CPU core and exclusively controlling a single n300 card.
%According to the results in Table~\ref{tab:times}, the Multi-Host Single-Chip strategy exhibits the best performance. Simulations performed on a single chip of an n300 card require, on average, $1459 \pm 0.47 , \mathrm{s}$, corroborating the $\approx 2\times$ speedup relative to CPU-only executions, as reported in~\cite{almerol2025SC}. When both chips on the same card are utilized, the execution time increases by $\approx 3.6 \%$. Even longer times-to-solution ($\approx 6.58 \times$ compared to the Single-Chip configuration) are observed when running the code on two chips and one card in the mesh-based configuration. Interestingly, when using two chips on two different cards through the Multi-Host Single-Chip approach, the simulation takes less time to complete with both respect to the usage of one card and the usage of one card, but two chips in the Multi-Host Multi-Chip approach.   

According to the results in Table~\ref{tab:times}, the Multi-Host Single-Chip strategy delivers the best performance. Simulations performed on a single chip of an n300 card require, on average, $1459 \pm 0.47 \, \mathrm{s}$ to complete, corroborating the $\approx 2\times$ speedup relative to CPU-only executions reported in~\cite{almerol2025SC}. For the CPU implementation, only the time-to-solution obtained using a shared-memory paradigm is reported, since this approach was verified to achieve superior performance on a single node compared to pure MPI or hybrid MPI+OpenMP configurations. When both chips on the same card are utilized, the execution time increases by approximately $3.6\%$. Even longer times-to-solution (about $6.58\times$ higher than the Single-Chip configuration) are observed when running the code on two chips and one card in the Mesh-Based configuration. This degradation can be attributed to the additional communication overhead introduced by the Ethernet interconnect between accelerators, which has lower bandwidth and higher latency compared to the PCIe link. Interestingly, when two chips located on different cards are employed through the Multi-Host Single-Chip approach, the simulation completes faster than when using a single card with two active chips, confirming that communication topology plays a crucial role in overall performance. In the Mesh-Based approach, data is distributed across devices using sharded buffers, with subsets of the global particle data assigned to each device. Unlike the first two approaches, which explicitly map data to Tensix cores, the mesh abstraction relies on runtime-managed buffer distribution and command scheduling. This simplifies multi-device programming but introduces overheads from buffer management, shard metadata handling, and synchronized command dispatch. In our experiments, we used device-level sharding to scale the $N$-body code across multiple devices. While this enables multi-chip execution, it does not fully exploit intra-device parallelism \citep{tenstorrent_tensor_sharding}. We are currently investigating strategies to optimize communication and data movement in the Multi-Host Multi-Chip configuration, as well as ways to better levarege the intra-device parallelism in the Mesh-Based approach to mitigate performance bottlenecks.
\begin{table}[t]
\centering
\resizebox{\columnwidth}{!}{
\begin{tabular}{| c | c | c | c | c | c |}
\hline
 Approach & \# cards & \# chips &\begin{tabular}[c]{@{}c@{}}Time-to-solution\\ {[}s{]}\end{tabular} & EDP {[}kJ s{]} \\ 
 \hline 
 \begin{tabular}[c]{@{}c@{}}Multi-Host\\Single-Chip\end{tabular}  & 1 & 1 & 1459.46 $\pm$ 0.47 & 563.10 $\pm$ 6.47\\ %19 sims
 \hline 
  \begin{tabular}[c]{@{}c@{}}Multi-Host\\Single-Chip\end{tabular} & 2 & 2 & 1318.54 $\pm$ 2.72 & 479.43 $\pm$ 5.41\\
\hline 
  \begin{tabular}[c]{@{}c@{}}Multi-Host\\Multi-Chip\end{tabular} & 1 & 2 & 1511.67 $\pm$ 0.95 & 636.84 $\pm$ 0.82\\
\hline 
  \begin{tabular}[c]{@{}c@{}}Mesh-Based\\Configuration\end{tabular} & 1 & 2 & 9614.54 $\pm$ 1.14 & 27822.11 $\pm$ 52.17\\ 
\hline 
\begin{tabular}[c]{@{}c@{}}CPU 32 OMP threads\\+ AVX-512\end{tabular} & 0 & 0 & 2875.39 $\pm$ 5.30 & 1927.70 $\pm$ 7.30\\
\hline
\end{tabular}
}
\caption{Execution time and EDP for simulations modeling 409600 particles over three time steps. Reported values correspond to the average time-to-solution and EDP over approximately 20 runs for the three approaches described in Section~\ref{sec3:porting}. The table also specifies the number of accelerator cards (second column) and processor chips (third column) used in each simulation, and includes, in the last row, the average execution time and EDP for CPU-only simulations. Note that simulations using a single n300 card employ one MPI rank, regardless of the number of chips utilized per card, whereas simulations using two cards and one chip per card in the Multi-Host Single-Chip approach employ two MPI ranks.} \label{tab:times}
\end{table}

\begin{figure}[]
\centering
\includegraphics[width=0.5\linewidth]{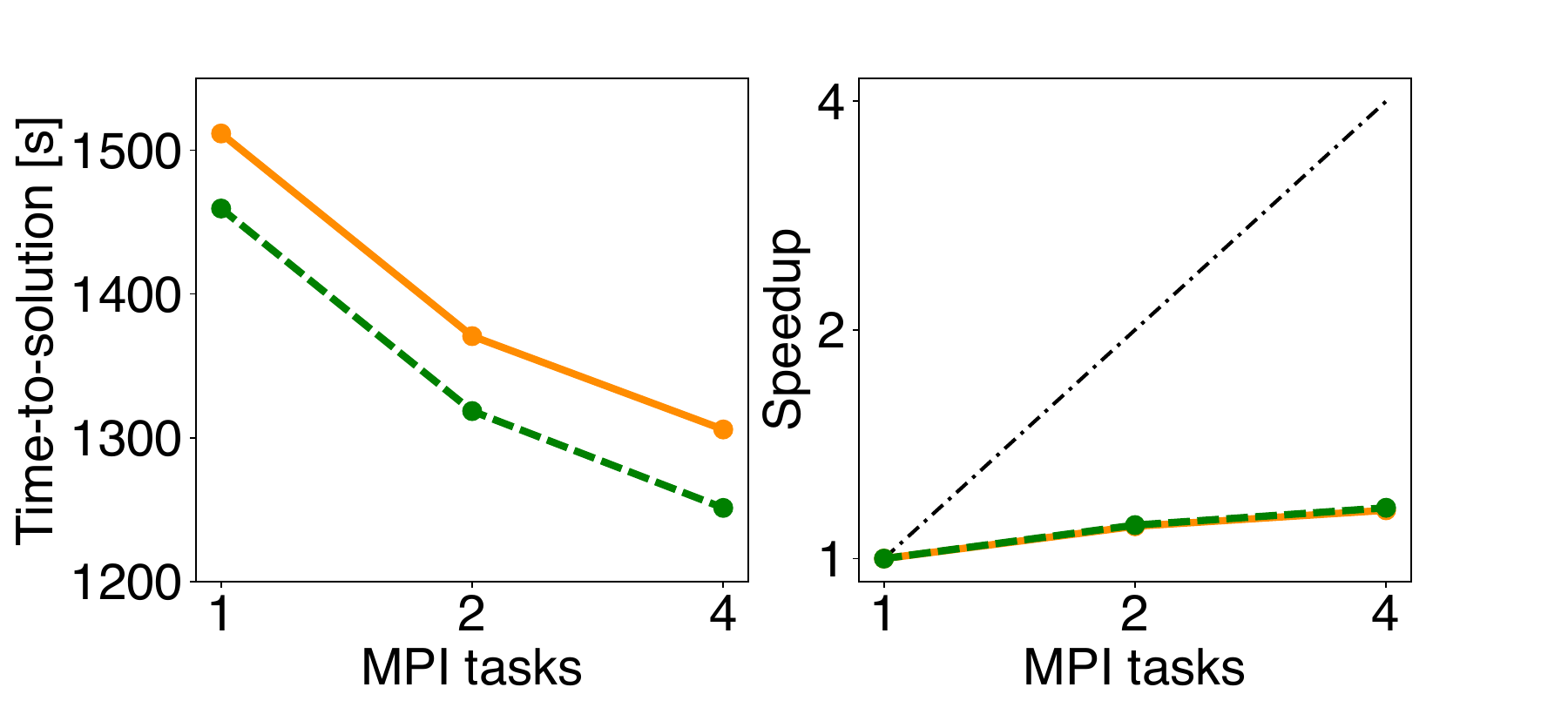}
\caption{Time-to-solution (left panel) and strong scaling parallel speedup (right panel) as functions of the number of MPI tasks for simulations performed using the Multi-Host Single-Chip (green dashed line) and Multi-Host Multi-Chip (orange solid line) configurations. The black dash-dotted line represents the ideal speedup.}\label{fig:strong_scaling}
\end{figure}

\paragraph{Scalability} We perform strong scaling tests for the Multi-Host Single-Chip and Multi-Host Multi-Chip approaches, which currently appear more promising than the Mesh-Based Configuration. The results are presented in Fig.~\ref{fig:strong_scaling}, which shows the decrease in time-to-solution with increasing number of MPI tasks (left panel) and the corresponding speedup (right panel). Reported values represent averages over at least 16 simulations. For the Multi-Host Single-Chip setup, the number of MPI tasks corresponds to the number of cards and chips used; for the Multi-Host Multi-Chip setup, the number of MPI tasks corresponds to the number of cards, while the number of chips is doubled. Shorter execution times are consistently achieved with the Multi-Host Single-Chip approach, which is on average $\approx 1.04\times$ faster for the same number of MPI tasks. Scalability appears comparable for both strategies, though far from ideal, with speedups and parallel efficiencies of $1.10\times$ and $55\%$, and $1.16\times$ and $29\%$ at two and four MPI tasks, respectively, for the Multi-Host Single-Chip configuration. These observations indicate that, for both configurations, further optimization of the communication and workload distribution strategies is likely necessary to achieve improved parallel performance.
We emphasize that the limited scalability is not caused by inter-card communication overhead, as Wormhole cards do not communicate directly with one another. Rather, the observed scaling limitations are likely due to overheads in the TT-Metalium multi-device orchestration layer, reflecting the still limited maturity of the software stack for multi-card execution.
Although our scalability tests were conducted on a single node hosting four cards, the same experiments could be performed on two or four host systems with two or one card each, respectively, without requiring code modifications. However, these configurations would likely exhibit lower scalability due to the higher latency of inter-node communication compared with the single-node setup used in our experiments.

Energy-to-solution values are obtained by combining raw and post-processed data sampled in user space at a frequency of $\approx 1 \, \mathrm{Hz}$ over the entire duration of the job, including both simulation and sleep periods~\citep{amati}. 

Following the methodology outlined in~\cite{almerol2025SC}, we record the timeseries of the power drawn by the eight chips on the four n300 cards, as well as the energy consumed by the two CPUs. For measuring the chip power, we use the Tenstorrent system management interface \verb|tt-smi|, while CPU energy is obtained via \verb|perf stat -a -e| with a one-second sampling interval.

The energy dissipated by a single n300 chip during a simulation is calculated as the discrete integral of its power over the simulation duration, excluding the \verb|sleep| periods. These values are then summed across all eight chips to obtain the total energy consumption of the four cards. Similarly, the energy consumed by the dual-socket processor is computed by summing the values recorded with \verb|perf|, considering only the active simulation time.

Conversely, given the energy recorded for each $1 \, \mathrm{s}$ interval, the instantaneous power absorbed by the CPUs is obtained as the discrete derivative of the energy.

The total energy per simulation (energy-to-solution) is then calculated as the sum of the energy consumed by both the four n300 cards and the dual CPUs. Given the energy-to-solution and the time-to-solution, it is also possible to compute the so-called Energy–Delay Product (EDP), defined as the product of the energy-to-solution and the time-to-solution~\citep{amati}. Table~\ref{tab:times} reports the EDP values for simulations performed using the three implemented approaches, as well as for the reference CPU simulation. The results show that the Multi-Host Single-Chip approach not only minimizes the time-to-solution, but also achieves the lowest EDP. Finally, to determine the overall power, we sum the power values from the n300 cards and the CPUs, performing linear interpolation when measurements are not aligned in time.

\begin{figure}[]
\centering
\includegraphics[width=0.5\linewidth]{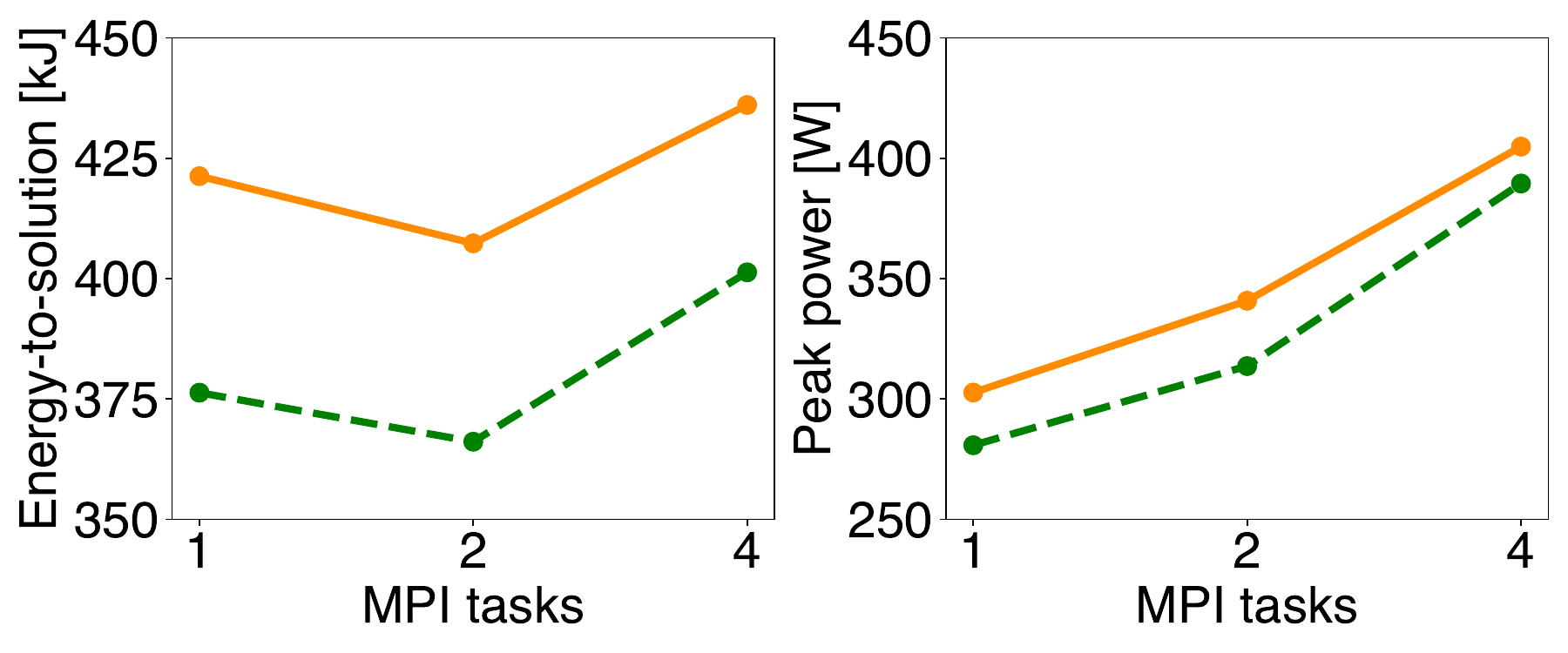}
\caption{Energy-to-solution (left panel) and peak power (right panel) as functions of the number of MPI tasks for simulations performed using the Multi-Host Single-Chip (green dashed line) and Multi-Host Multi-Chip (orange solid line) configurations.}\label{fig:energy_and_power}
\end{figure}

The left panel of Fig.~\ref{fig:energy_and_power} presents the energy-to-solution as a function of the number of MPI tasks for the Multi-Host Single-Chip and Multi-Host Multi-Chip configurations. As for the time-to-solution, these values represent averages over multiple simulation runs. Unlike the time-to-solution, however, the energy-to-solution does not exhibit a monotonic trend, displaying instead a minimum when two MPI tasks are employed. In particular, using two MPI tasks yields an energy reduction of approximately $10 \%$ compared to the configuration achieving the shortest time-to-solution in both cases. The EDP results minimized in this case. This indicates that running with two MPI tasks provides a good compromise between execution time and overall energy efficiency.

The right panel of Fig.~\ref{fig:energy_and_power} reports the maximum system peak power observed (excluding memory, storage, networking, cooling system, and other contributions) for simulations using one, two, and four MPI ranks. Employing two chips per card increases the peak power by up to $\approx 8.7\%$ relative to configurations using a single chip per card.

\section{Summary and perspectives}\label{sec:conclusion}
Recently, we developed an $N$-body astrophysical simulation code and accelerated its force calculation using the Tenstorrent TT-Metal programming interface. By executing the code on the RISC-V-based Tenstorrent Wormhole n300 accelerator, we demonstrated that this class of devices can offer a competitive and energy-efficient solution for next-generation astrophysical computations.

In the present work, we further investigated methodologies to enhance code parallelism. In particular, we enabled execution across multiple cards simultaneously by exploiting MPI and card-level parallelism. To the best of our knowledge, this represents the first example of a scientific code capable of running across multiple Tenstorrent devices and cards.

Three approaches were implemented to scale simulations across cards and devices: the Multi-Host Single-Chip Configuration, the Multi-Host Multi-Chip Configuration, and the Mesh-Based Configuration. Among these, the first approach delivered the best performance in terms of both time-to-solution and energy-to-solution, while the Mesh-Based Configuration yielded the lowest performance. In terms of scalability, the first and second approaches showed comparable behavior, both exhibiting relatively low parallel efficiency at four MPI tasks, suggesting that further optimization is required. Our results also show that while the time-to-solution decreases monotonically with the number of MPI tasks, the energy-to-solution and the enrgy delay product display a minimum when using two MPI tasks, indicating that this configuration provides an effective trade-off between execution time and energy efficiency.

As an immediate next step, we plan to benchmark the Tenstorrent results against conventional NVIDIA and AMD GPUs, as well as Xilinx Alveo FPGAs. We have already started developing the force kernel in CUDA. Preliminary results obtained on a system featuring an AMD Genoa CPU and an NVIDIA H100 GPU, using the same parameters described in Section~\ref{sec:results}, show a speedup of $2.6\times$ with respect to the Multi-Host Single-Chip approach. These results indicate a performance gap with respect to more established accelerator technologies, which can reasonably be attributed to the relative maturity of the current RISC-V–based accelerator platform and its software stack. Further evaluation on newer hardware is therefore required to better assess the evolution of performance, and for this reason we plan to extend our experiments to the Tenstorrent p150 (codename Blackhole\texttrademark).
Since the p150 is built upon the same underlying architecture as the Wormhole series, we anticipate that the strategies developed in this work will transfer effectively to this platform.

\section*{Acknowledgements}
We thank E. Duffy, R. Friedman, R. Ganisetti, and F. LeClair (Tenstorrent) for valuable discussions.

This research is supported by the Italian Research Center on High Performance Computing Big Data and Quantum Computing (ICSC), project funded by European Union - NextGenerationEU - and National Recovery and Resilience Plan (NRRP) - Mission 4 Component 2 within the activities of Spoke 3 (Astrophysics and Cosmos Observations).

E.B. and D.G. acknowledge support from the SPACE project, funded by the European Union. This project has received funding from the European High Performance Computing Joint Undertaking (JU) and from Belgium, the Czech Republic, France, Germany, Greece, Italy, Norway, and Spain under grant agreement No.~101093441.

\section*{Declaration of generative AI and AI-assisted technologies in the manuscript preparation process}
During the preparation of this work we used ChatGPT and Grammarly in order to improve the clarity and correctness of the English language. The content of the paper remains entirely original. After using these tools, we reviewed and edited the content as needed. We take full responsibility for the content of the published article.

\appendix
\section{Acceleration and Jerk Calculations within the Compute Kernel}
\label{app1}
\begin{algorithm}[t]
\caption{Tile-based pairwise acceleration and jerk evaluation within the Compute Kernel}
\label{alg:async_core}
\begin{algorithmic}[1]
\State Acquire lock on dst register
\State Compute relative displacement $\mathbf{r}_{ij} = \mathbf{r}_j - \mathbf{r}_i$
\State Store components $r_{x,ij}, r_{y,ij}, r_{z,ij}$ in intermediate circular buffers (CBs) for reuse
\State Acquire lock on dst register
\State Compute distance metrics $r_{ij}^2 = |\mathbf{r}_{ij}|^2$, $r_{ij}^{-1} = 1 / \sqrt{r_{ij}^2}$, and $r_{ij}^{-3} = (r_{ij}^{-1})^3$
\State Compute scaling coefficients $t_j = p_{w,j} \, r_{ij}^{-3}$ and $A_{ij} = -3 \, r_{ij}^{-2}$
\State Compute velocity difference $\mathbf{v}_{ij} = \mathbf{v}_j - \mathbf{v}_i$
\State Compute radial velocity component $v_{r,ij} = \mathbf{r}_{ij} \cdot \mathbf{v}_{ij}$
\State Compute scaling coefficient $q_{ij} = A_{ij} \, v_{r,ij}$
\State Store $t_j$ and $q_{ij}$ in CBs for reuse
\State Acquire lock on dst register
\State Update temporary acceleratio $
\mathbf{a}_i^{\,t} \mathrel{+}= t_j \, \mathbf{r}_{ij}
$
\State Acquire lock on dst register
\State Update temporary jerk
$\dot{\mathbf{a}}_i^{\,t} \mathrel{+}= t_j \, (\mathbf{v}_{ij} + q_{ij} \, \mathbf{r}_{ij})$
\end{algorithmic}
\end{algorithm}

Algorithm \ref{alg:async_core} outlines the tile-based pairwise acceleration and jerk computation implemented within the compute kernel of the Tenstorrent Wormhole accelerator.
Intermediate quantities such as the relative displacement, distance metrics, and scaling coefficients are staged into circular buffers (CBs) to enable data reuse and minimize redundant computation across tiles.

Synchronization primitives are used to acquire and release locks on destination registers, ensuring safe sequencing between computation and data movement stages. Releasing the destination register before staging data to a CB allows other compute or DMA operations to proceed concurrently, improving pipeline utilization.

A small softening parameter, $\epsilon = 1.0 \times 10^{-7}$, is included in the distance computation to avoid numerical singularities when particles are close.

%%       \citet{<label>} ==> Lamport (1994)
%%       \citep{<label>} ==> (Lamport, 1994)
%%
% Example citation, See \citet{lamport94}.




\bibliographystyle{elsarticle-harv} 
\bibliography{main}

\end{document}